\documentclass[twocolumn,showpacs,preprintnumbers,amsmath,amssymb]{revtex4} 
\usepackage{graphicx}% Include figure files 
\usepackage{dcolumn}% Align table columns on decimal point 
\usepackage{bm}% bold math 
\usepackage{color}

\newcommand{\bx}{{\bm x}}

\begin{document} 
 
%\preprint{For submission to {\it Physical Review Letters}} 
 
\title{Spatially Localized Quasicrystals}% Force line breaks with \\ 
 
\author{P.~Subramanian$^1$, A.J.~Archer$^2$, E.~Knobloch$^3$ and A.M.~Rucklidge,$^1$}
\affiliation{$^1$Department of Applied Mathematics, University of Leeds, Leeds LS2 9JT, UK\\
$^2$Department of Mathematical Sciences, Loughborough University, Loughborough LE11 3TU, UK\\
$^3$Department of Physics, University of California at Berkeley, Berkeley CA 94720, USA}

\date{\today}% It is always \today, today, 

\begin{abstract}

We investigate quasicrystal-forming soft matter using a two-scale phase field
crystal model. At state points near thermodynamic coexistence between
bulk quasicrystals and the liquid phase, we find multiple metastable
spatially localized quasicrystals embedded in a background of liquid. 
In three dimensions we obtain spatially localized icosahedral quasicrystals.
In two dimensions, we compute several families of spatially
localized quasicrystals with dodecagonal structure and investigate
their properties as a function of the system parameters. In both cases the
localized quasicrystals are metastable, and so correspond to energetically
locally favored structures. The presence of such structures is expected
to crucially affect the dynamics of the crystallization process.

\end{abstract}
 
\pacs{61.44.Br, 81.10.Aj, 61.50.Ah}% PACS, the Physics and Astronomy 

% PACS numbers:
% 61.44.-n -- Semi-periodic solids
% 81.10.Aj -- Theory and models of crystal growth; physics and chemistry of crystal growth, crystal morphology, and orientation 
%             also nucleation
% 61.50.Ah -- Theory of crystal structure, crystal symmetry; calculations and modeling  
% 05.20.-y -- Classical statistical mechanics -- no
% 61.44.Br -- Quasicrystals
% 64.70.D  -- Solid-liquid transitions

% \keywords{Suggested keywords}%Use showkeys class option if keyword 

\maketitle 
%\section{Introduction}

Soft matter is capable of organizing into complex structures even when the underlying
interactions between the individual constituents are quite simple. These structures can be described by the phase field crystal (PFC) model\hbox{\,\cite{Emmerich2012}}.
This model represents a simplification\hbox{~\cite{van2009derivation, archer2012solidification}} of the more accurate
Density Functional Theory (DFT) approach\hbox{~\cite{Evans1992,Hansen2013}}
and has proved useful in understanding the formation and properties of soft matter crystals
under various conditions. 
The simplest such model takes the form of the conserved Swift--Hohenberg
equation and the steady states of this model have been studied in detail
in one, two and three dimensions \cite{Thiele2013}. The model
exhibits coexistence between the liquid state and a crystalline state over a 
range of temperatures \cite{Emmerich2012, elder2004modeling}.
In this regime one also finds spatially localized structures that bifurcate
from the periodic crystal close to the spinodal, and exhibit homoclinic snaking
\cite{Knobloch2016,Thiele2013} before reconnecting
to the periodic state. In finite size systems, these localized
structures can have lower energy than the periodic state and so be
thermodynamically preferred.
%(see localization in infinite domains discussion by Edgar in appendix)}.

The PFC model, suitably modified to allow interaction between distinct spatial
scales~\cite{Engel2007,Subramanian2016}, also forms spatially extended quasicrystals
(QCs). Both dodecagonal QCs in two dimensions (2D)~\cite{Achim2014} and
icosahedral QCs in three dimensions (3D)~\cite{Subramanian2016} have been
identified and these may be thermodynamically stable under appropriate
conditions. Significantly, these extended QC states can arise at state points
where the uniform liquid state is metastable, in the so-called subcritical regime. This fact
raises the possibility that stable localized structures with a QC motif are also
present in this region \cite{Knobloch2016}. Here we show that such
structures are indeed present and investigate their stability properties. Of
course, a localized structure cannot, strictly speaking, be a quasicrystal. In
the following we use this terminology to refer to structures that turn into an
extended QC in a continuous manner as one follows the localized solution, varying
one of the system parameters. Loosely speaking, we may think of such localized structures as a
QC state with a superposed envelope that locally picks out a particular motif
from the QC state and suppresses the pattern further away.

%\section{Model}

Our model describes the space-time
evolution of a dimensionless scalar field $U(\bx,t)$ that specifies the
location and magnitude of density perturbations from a homogeneous
state~\cite{Emmerich2012}. The system is described by the Helmholtz free
energy%~$\mathcal{F}$
 \begin{equation}
 {\mathcal F}\left[U\right]=\int \Big[
           -\frac{1}{2}U{\mathcal L}U - \frac{Q}{3}U^3 + \frac{1}{4}U^4
           \Big]d^d\bx.
 \label{eq:fe}
 \end{equation}
 The evolution of $U$ conserves mass, i.e.\ ${\bar U}\equiv\int_DU\,d^d{\bf x}$ is fixed, where $D$ is the system volume, and follows
 \begin{equation}
 \frac{\partial U}{\partial t}={\nabla^2}\left(\frac{\delta \mathcal{F}[U]}{\delta U}\right)=-{\nabla^2}\left(\mathcal{L}U+QU^2-U^3\right).\label{eqn2}
 \end{equation}
Here the linear operator $\mathcal{L}$, chosen as in
\cite{RucklidgePRL2012,Subramanian2016}, allows for the growth of density modulations at two
wave numbers $k=1$ and $k=q<1$. The growth rate $\sigma(k)$ of a mode with wave number~$k$ is
given by a fifth order polynomial in $k^2$~\cite{Subramanian2016} that is
characterized by the ratio $q$ of the two wave numbers, two parameters $r_1$ and
$r_q$ that control the linear growth rates of the modes $k=1$ and $k=q$, and by a negative parameter~$\sigma_0$ that
controls the sharpness of the peaks in the growth rate at these two wave
numbers. Our model allows independent control of these three quantities, in
contrast to models based on the Lifshitz--Petrich free energy~\cite{Lifshitz1997,Achim2014}.
In particular, when $r_1=r_q={\bar U}=0$, modes with wave number~$k\neq q,1$ decay as
$\sigma(k)=\sigma_0 k^2(1-k^2)^2\,(q^2-k^2)^2/q^4$. When $r_1\neq0$ or $r_q\neq0$
there are additional terms in $\sigma(k)$ and when ${\bar U}\neq 0$ the growth rate
$\sigma(k)$ is also influenced by the parameter $Q$~\cite{Subramanian2016}. %{\bf correct?}
The parameter $Q$ also controls the strength of three-wave
interactions. With appropriate choices of $q$, such as
$q=1/\tau_2$ with $\tau_2=2\cos(\pi/12)$ or $q=1/\tau_3$ with
$\tau_3=2\cos(\pi/5)$, we promote the formation of dodecagonal QCs in 2D
($d=2$) or icosahedral QCs in 3D ($d=3$)~\cite{Archer2015,Subramanian2016}.

The equilibrium states of the dynamical system (\ref{eqn2}) solve
${\partial U}/{\partial t}=0$, and hence satisfy
 \begin{equation}
 \mathcal{L}U+QU^2-U^3=\textrm{constant}=-\mu\,,\label{eqn3}
 \end{equation}
where $\mu\equiv\delta {\mathcal F}/\delta U$ is the chemical potential. 
We characterize computed structures in terms of the grand (Landau) potential
 \begin{equation}
 \Omega[U] = {\mathcal F}[U] - \mu\int U d^d\textbf{x}\,,
 \end{equation}
whose minima correspond to metastable states. The global minimum amongst all 
states corresponds to
the thermodynamically stable state. All results presented here are computed at~$\mu=0$,
so states that minimize~$\Omega$ also minimize~$\mathcal{F}$.
For a given value of~$\mu$, there can be several possible bulk extended 
equilibria with different values of~${\bar U}$; we focus on the one with 
lowest~$\Omega$. With $\mu=0$, the only 
homogeneous equilibrium for our parameter values is the uniform liquid with
${\bar U}=0$, in which case $\sigma(1)=r_1$ and $\sigma(q)=q^2r_q$.

\begin{figure}
\centering
\includegraphics[width=.95\hsize]{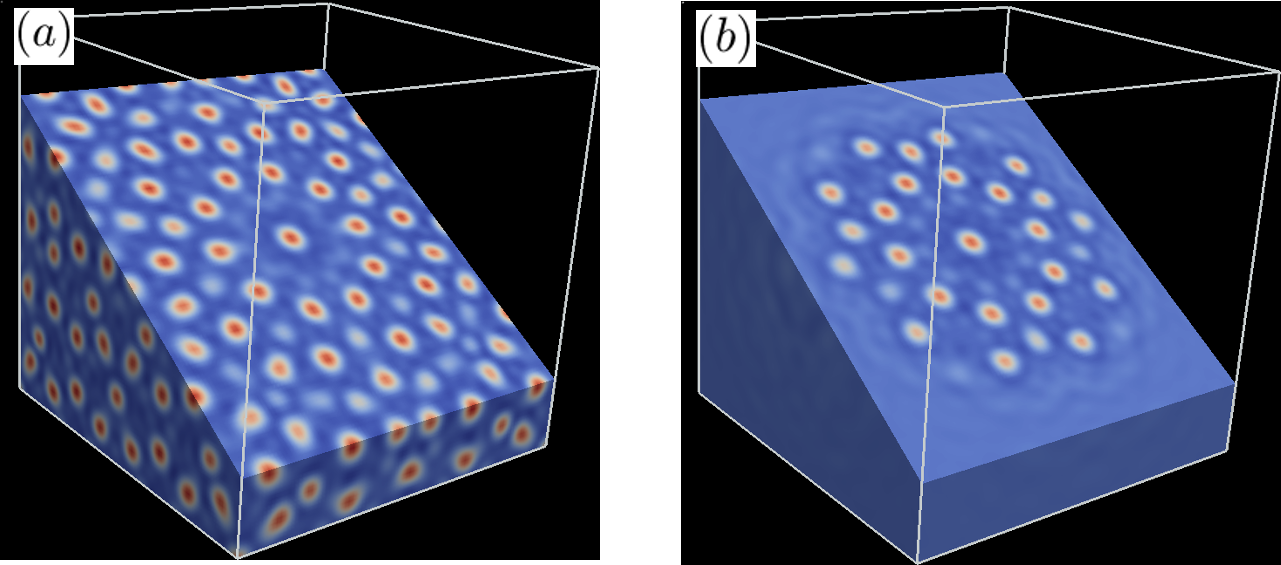}
\caption{The 3D scalar density $U({\bf x})$ sliced along the plane
($\tau_3,0,1$)  for ($a$) a thermodynamically stable extended icosahedral QC
and ($b$) a metastable localized quasicrystal at the same parameters, with
$r_1=r_q=-0.51$, $q=0.6180=1/\tau_3$, $Q=2$, $\sigma_0=-10$ and $\mu=0$. See \cite{Supplement_SLQ} for more details on the structure of the extended and spatially localized QCs.\label{fig1:3Dlocal}}
 \end{figure}

In~\cite{Subramanian2016}, we showed that minima of~$\mathcal{F}$ in 3D domains
correspond to periodic structures such as lamellae, columnar hexagons, and
body-centered cubic crystalline phases, as well as an icosahedral
QC such as that shown in Fig.~\ref{fig1:3Dlocal}($a$). For
parameter values where the homogeneous liquid phase and the icosahedral
QC state are both linearly stable, we also find states
that minimize~$\Omega$ and correspond to \emph{spatially localized quasicrystals},
i.e., states in which the QC phase fills only part of the domain. 
Figure~\ref{fig1:3Dlocal}($b$) shows one such state. This state retains icosahedral
symmetry and the slice is chosen to reveal its approximate 5-fold symmetry. Such spatially
localized equilibria lie on a continuous solution branch when a system parameter
such as $r_1$ varies, and we find that this branch connects with the branch of fully extended QCs.
Since the numerical continuation of branches of 3D localized solutions involves
challenging computations we focus in the rest of this Letter on localized
dodecagonal QCs in 2D.

%%%%%%%%%%%%%%%%%%%%%%%%%%%%%%%%%%%%%%%%%%%%%%%%%%%%%%%%%%%%%%%%%%%%%%%%%%%%%%

%\section{Numerical Details}

Two-dimensional direct numerical simulations of Eq.~(\ref{eqn2}) are
carried out in a periodic square domain of side length $30\times 2\pi$. 
This choice of the domain size allows the formation of
extended (but still approximate)~\hbox{QCs¬\cite{Rucklidge2009}}. In a pseudospectral approach we
employ 256 Fourier modes using FFTW \cite{FFTW05} in each direction
and time-step using second-order exponential time differencing (ETD2)
\cite{CoxJCP2002}.
Starting from smoothed random initial conditions, we find several
qualitatively distinct equilibria: the liquid
state, lamellae (lam) and hexagons (hex) at each of the two
possible wavelengths ($2\pi$ and $2\pi/q$), and a dodecagonal~\hbox{QC}.
These equilibrium solutions are then continued numerically using 
pseudo-arclength continuation~\cite{Doedel1991} employing the
biconjugate gradient stabilized method~\cite{vanderVorst1992} or with the
induced dimension reduction method \cite{Sonneveld2008} to solve the 
large linear systems that arise in the
Newton iteration up to an accuracy of $\mathcal{O}(10^{-10})$. Numerical continuation allows us to
explore the space of possible fully nonlinear equilibria of the
model irrespective of their dynamical stability.

%\section{Results}

\begin{figure}
\hspace{-0.2cm}
 {\includegraphics[width=0.99\hsize]{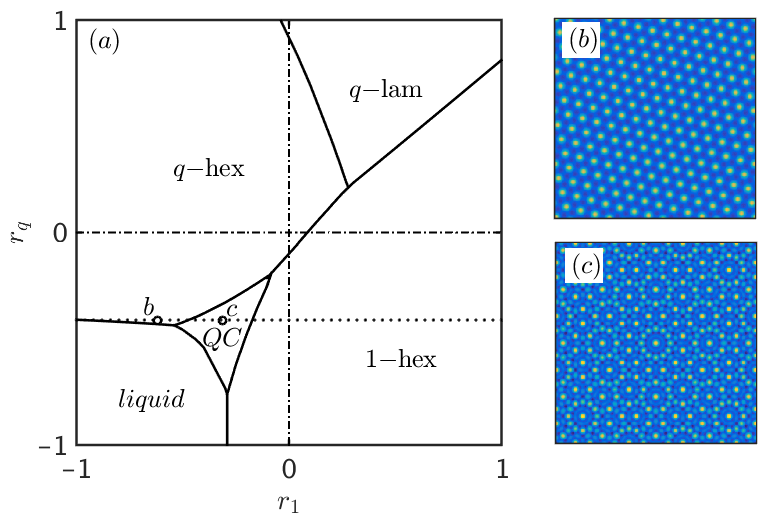}}
\vspace{-2.5ex}
 \caption{($a$) Thermodynamically stable 2D structures in the $(r_1,r_q)$ plane,
   computed as global minima of the grand potential $\Omega$ when $\mu=0$.
   Dashed-dotted lines indicate the axes. The dotted line represents a cut
   corresponding to $r_q=-0.412$ explored in detail in subsequent plots.
   Panels ($b$) and ($c$) show stable states along this cut, $q$-hex at $r_1=-0.6181$ and QC at
$r_1=-0.3112$, respectively. The remaining parameters are: $q=0.5176=1/\tau_2$, $Q=2$ and
$\sigma_0=-10$.}
 \label{fig2:phaseplot}
 \end{figure}

%\subsection{phase plot and Maxwell points}

Figure~\ref{fig2:phaseplot}($a$) shows the $(r_1,r_q)$ parameter plane for
chemical potential $\mu=0$, indicating the global minimum of the grand
potential~$\Omega$ at each combination of parameter values. When both $r_1$ and
$r_q$ are strongly negative, the liquid state is the global minimum. When $r_1$
or $r_q$ increase we obtain hexagons with lattice spacing $2\pi$ or $2\pi/q$
[$1$-hex or $q$-hex, Fig.~\ref{fig2:phaseplot}($b$)] and a
region of $q$-lamellae when both are positive. Of particular
interest is a substantial region in the third quadrant with stable dodecagonal
QCs [Fig.~\ref{fig2:phaseplot}($c$)]. In this region the linear growth rates of
the $k=1$ and $k=q$ modes are both negative, implying the uniform liquid is
linearly stable, albeit metastable with respect to the QC. Parameter values
where the grand potentials of two different states are equal, i.e., the Maxwell
points, separate the different regions [solid lines in
Fig.~\ref{fig2:phaseplot}($a$)]. Maxwell points are ideal starting
parameter combinations for locating spatially localized states consisting of a
patch of one state embedded in the background of another.

%%%%%%%%%%%%%%%%%%%%%%%%%%%%%%%%%%%%%%%%%%%%%%%%%%%%%%%%%%%%%%%%%%%%%%%%%%%%%%%%%%
%\subsection{bifurcation plot and extended state branches}

\begin{figure*}[]
{\includegraphics[width=0.95\textwidth]{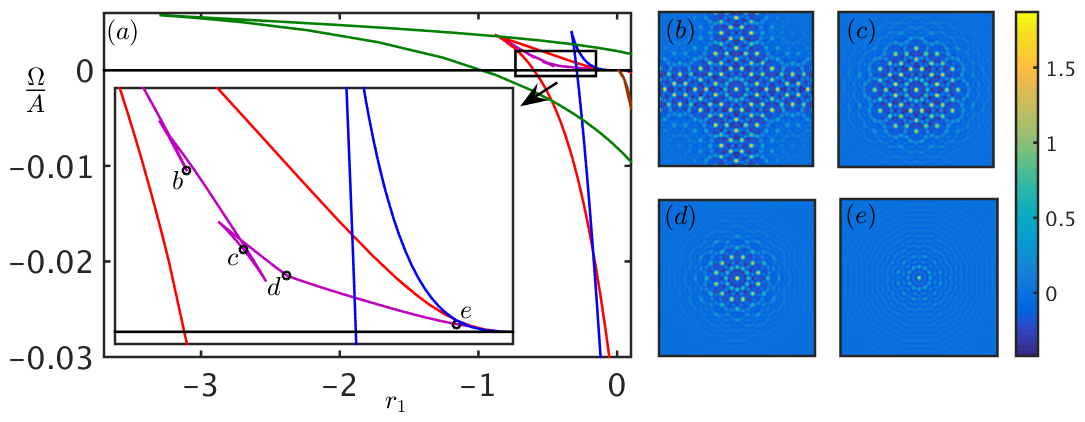}}
\vspace{-2ex}
\caption{($a$)~Specific grand potential $\Omega/A$ ($A$ is the area of the 
domain) as a function
of~$r_1$ [dotted line in Fig.~\ref{fig2:phaseplot}($a$)], with $\mu=0$ and
$r_q=-0.412$. The blue line indicates $1$-hexagons, 
the green line indicates $q$-hexagons, the red
line indicates the extended QC state, while the magenta
line indicates the branch of spatially localized QC states with dodecagonal
symmetry.  ($b$)-($d$)~Metastable localized QC states on each `swallow-tail'
at locations indicated in the inset in panel ($a$) corresponding to
$r_1=-0.5954$, $-0.4926$ and $-0.4151$, respectively. ($e$)~Unstable
localized state at $r_1=-0.14$. The color scale indicates the
value of $U({\bf x})$ in panels ($b$)-($e$). The remaining parameters are as
in Fig.~\ref{fig2:phaseplot}.}
 \label{fig3:bifplot}
 \end{figure*}

In order to investigate the effect of varying $r_1$, we set $r_q=-0.412$ and
$\mu=0$ [dotted line in Fig.~\ref{fig2:phaseplot}($a$)] and study the
bifurcation behavior of the system. Figure~\ref{fig3:bifplot}($a$) shows the
specific grand potential as a function of~$r_1$ for spatially extended
states: liquid (black line), $1$-lam (brown line), $1$-hex (blue line), $q$-hex
(green line) and the dodecagonal QC (red line). The $1$-lam, $1$-hex and QC
structures originate at the spinodal (threshold for linear instability of the liquid state) point $r_1=0$.
The hexagonal and the QC
states arise via transcritical bifurcations and so are found on either side of
$r_1=0$. However, linearly stable states are only found along parts of the branch below the
prominent cusps.

%\subsection{snaking branch}

The inset in Fig.~\ref{fig3:bifplot}($a$) shows the vicinity of the coexistence between
the QC states (red line) and the liquid (black line). The QC and liquid
are in thermodynamic coexistence, at the crossing point of these lines at $r_1=-0.6$. 
Spatially localized QC states, shown in magenta, are found nearby. Panels
($b$)--($e$) show sample solutions along this branch at the locations indicated in the
inset of panel ($a$). These states arise via spatial modulations of Eckhaus type, i.e., long
wavelength modulations that suppress the QC amplitude in different parts of the
spatial domain \cite{Bergeon2008}. With increasing $r_1$, the branch of
unstable localized states undergoes fold bifurcations creating three
`swallow-tails' close to locations ($b$), ($c$) and ($d$), before connecting to the extended QC branch
close to the QC spinodal, near location~($e$).

Panel ($b$) shows an example of a spatially modulated QC, where the QC
structure is suppressed in the vicinity of the four corners of the domain. The
state that remains is metastable 
and is organized around a dodecagonal structure at its
center. As one follows the branch of modulated QC states (magenta line) to the
next swallow-tail, these holes deepen, suppressing the QC structure further 
and leaving
a spatially localized QC in a background liquid state. Panel ($c$) shows a
metastable state of this type. As $r_1$ increases from the left fold close
to ($c$), the structure starts to lose the outer ring of 12 peaks. This process
is almost complete by location ($d$), where there is only one ring of 12 peaks
surrounding a single central peak. This state is also metastable. With
further increase in $r_1$, the state gradually decreases in amplitude, while
the interface becomes more diffuse, as the spatially modulated structure
approaches an unstable low amplitude but spatially extended QC solution 
near the spinodal. Panel~($e$) shows an
example of an unstable localized state just prior to the merger with
the unstable extended QC state at $r_1=-0.11$.

Metastable localized states are found only on the lower branch in each of the
three swallow-tail regions, and these regions are a reflection of slanted
snaking that is expected of localized structures in mass-conserving
systems~\cite{Dawes2008,Thiele2013,Knobloch2016}. We anticipate that in larger domains the
number of swallow-tails will be greater since each is associated with the appearance
or disappearance of a layer of structure around the central dodecagonal motif.
Similar spatially localized states associated with the other crystalline structures
shown in Fig.~\ref{fig2:phaseplot}, e.g.,
hexagon patches and target patterns, are also expected to be present, as is the
case in non-conserved systems~\cite{Lloyd2008}.

%%%%%%%%%%%%%%%%%%%%%%%%%%%%%%%%%%%%%%%%%%%%%%%%%%%%%%%%%%%%%%%%%%%%%%%%%%%%%%%%%%%
% \subsection{symmetry broken localized states}

\begin{figure*}
{\includegraphics[width=0.95\hsize]{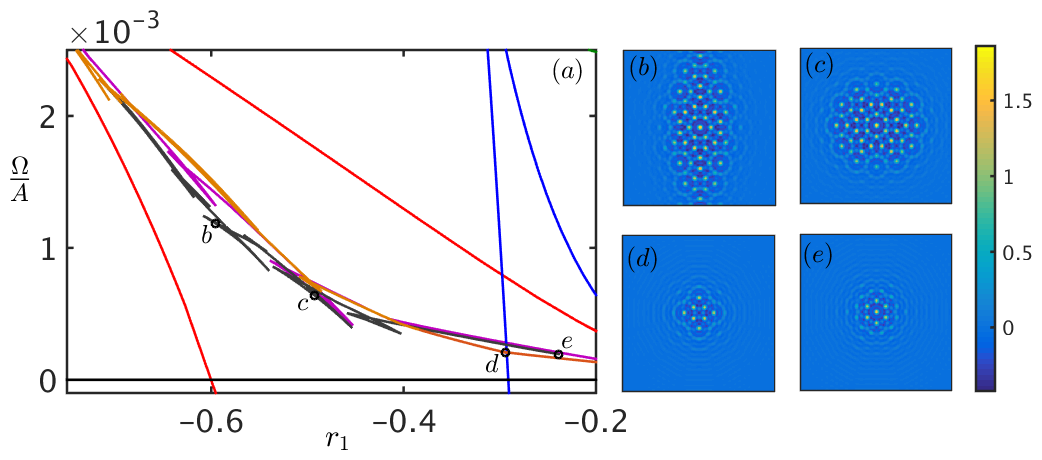}}
\vspace{-2ex}
\caption{($a$) Specific grand potential as a function
of~$r_1$. In addition to lines from the inset in Fig.~\ref{fig3:bifplot}($a$),
the dark gray line is a branch of localized states with
hexagonal central motif and the golden line is a branch of states 
with square central motif. The corresponding profiles $U({\bf x})$~at ($b$) $r_1=-0.5954$, 
($c$) $r_1=-0.4926$, ($d$) $r_1=-0.2937$ and ($e$) $r_1=-0.2388$. 
The remaining parameters are as in Fig.~\ref{fig2:phaseplot}.}
 \label{fig4:bifplotasymloc}
 \end{figure*} 

The dodecagonal symmetry about the central peak visible in
Fig.~\ref{fig3:bifplot} panels ($b$)--($e$) is retained all along the branch
of localized QC shown in magenta in Fig.~\ref{fig3:bifplot}($a$). Additional
branches of localized QC with reduced symmetry are also present.
Figures~\ref{fig4:bifplotasymloc}($d$,$e$) show two such equilibria with square
and hexagonal central motifs, for $r_1$ values near where the associated
localized states merge with the unstable extended QC state. In general
the growth of these structures with decreasing $r_1$ does not maintain the
symmetry of the central motif -- these states are, after all, quasicrystalline
and hence aperiodic -- unless enforced by the shape of the domain. We show
examples of equilibria that have broken the (approximate) hexagonal symmetry
of the central motif in Figs.~\ref{fig4:bifplotasymloc}($b$,$c$).

The branches of localized solutions associated with each of the symmetry-broken
motifs also display swallow-tails, as shown in Fig.~\ref{fig4:bifplotasymloc}($a$).
Localized solutions with square symmetry are shown in gold while states with
(approximate) hexagonal symmetry are shown in dark gray. The numbers and parameter
ranges of the swallow-tails along each branch differ. Among the branches of localized
solutions, the branch with hexagonal motif has the lowest grand potential over
most of its range.

%%%%%%%%%%%%%%%%%%%%%%%%%%%%%%%%%%%%%%%%%%%%%%%%%%%%%%%%%%%%%%%%%%%%%%%%%%%%%%%%

%\section{PFC vs. DFT result}

\begin{figure}
\includegraphics[width=0.95\hsize]{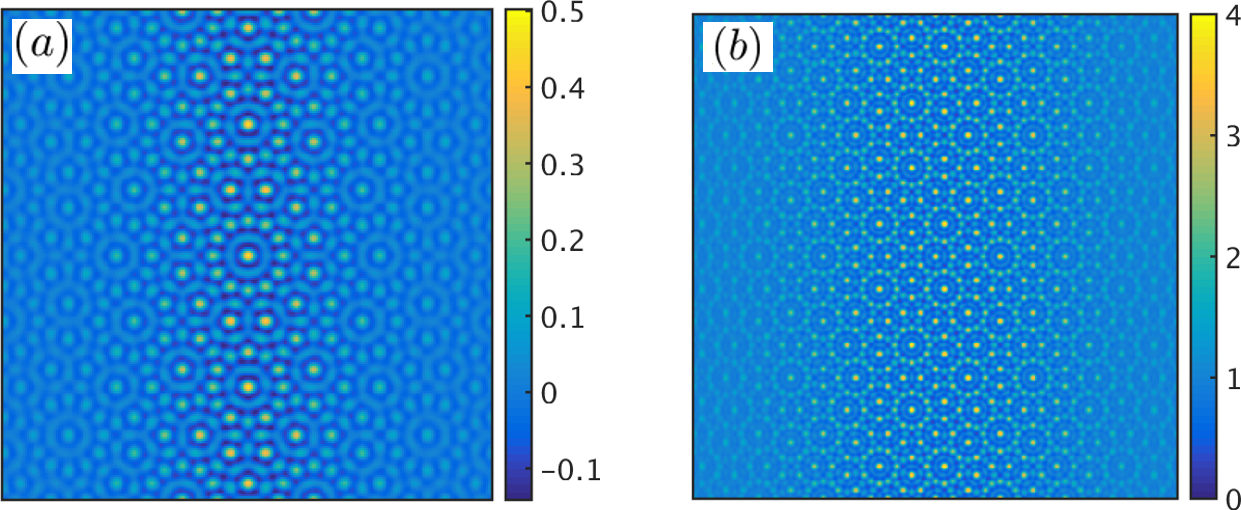}
\caption{($a$) Profile $U({\bf x})$, showing a 1D localized
QC in the PFC model at the parameter values in 
Fig.~\ref{fig2:phaseplot} and $r_1=-0.1152$. ($b$) Density profile obtained from DFT
for a system of particles interacting via the pair
potential \hbox{$v(r) = 10\,e^{-0.297r^2}
\left(1-1.09r^2+0.4397r^4-0.05r^6+0.002r^8\right)$} \cite{Barkan2014}, showing a 1D localized
QC similar to ($a$).}
 \label{fig5:DFTandPFC}
 \end{figure} 

Figure~\ref{fig5:DFTandPFC}($a$) shows $U({\bf x})$ for a localized QC state with
large scale modulation in one direction suppressing
the QC amplitude at either side of the domain. This coexistence of the liquid and
the QC has a rather diffuse interface between the
two states. In order to check the robustness of
spatial localization of QCs across
models, we compare this state with a similar equilibrium obtained from
DFT for a model system of particles with soft pair interactions
(following~\cite{Barkan2014}). For the chosen
parameters, we observe similar diffuse 1D localization of a dodecagonal
QC in a background liquid state [Fig.~\ref{fig5:DFTandPFC}($b$)].
In this example, the PFC solution is unstable, while the DFT localized state is metastable. We have not yet found a QC localized in 2D in the DFT model.

%%%%%%%%%%%%%%%%%%%%%%%%%%%%%%%%%%%%%%%%%%%%%%%%%%%%%%%%%%%%%%%%%%%%%%%%%%%%%%%%%

% discussion and future
 
We have shown that localized patches of QCs surrounded by bulk liquid can be
metastable both in 2D and 3D. In 2D these structures are located on the swallow-tails in 
Figs.~\ref{fig3:bifplot} and~\ref{fig4:bifplotasymloc}.
The fact that these spatially localized QCs are metastable has
significant consequences: the stability makes these structures longer lived
in an environment with thermal fluctuations and so will influence the rate of
QC nucleation. Moreover, these localized states will be
energetically locally favored structures (LFS)
\cite{tanaka1999two, mossa2003locally, Royall2008}. The fact that these LFS have QC
structure means that they will suppress the formation of the regular hexagonal
crystal. In 3D, icosahedral LFS are believed to be linked to dynamic arrest
\cite{Royall2008}, i.e., the avoidance of crystallization. Indeed, based on
the large spatial extent of some of the stable localized structures we find,
we conjecture that quite large LFS may in general be present in systems of this kind.

%\begin{acknowledgments} 

We are grateful to Laurette Tuckerman for assistance and discussions about the 
biconjugate gradient method. This work was supported in part by the L'Or{\'e}al UK \hbox{and Ireland Fellowship For Women} In Science (PS) and the EPSRC under
grants EP/P015689/1 (AJA) and \hbox{EP/P015611/1 (AMR)} and the NSF under grant
\hbox{DMS-1613132 (EK)}.

 %\end{acknowledgments}

%\bibliographystyle{apsrev} 
%\bibliography{literat}

\begin{thebibliography}{28}
\expandafter\ifx\csname natexlab\endcsname\relax\def\natexlab#1{#1}\fi
\expandafter\ifx\csname bibnamefont\endcsname\relax
  \def\bibnamefont#1{#1}\fi
\expandafter\ifx\csname bibfnamefont\endcsname\relax
  \def\bibfnamefont#1{#1}\fi
\expandafter\ifx\csname citenamefont\endcsname\relax
  \def\citenamefont#1{#1}\fi
\expandafter\ifx\csname url\endcsname\relax
  \def\url#1{\texttt{#1}}\fi
\expandafter\ifx\csname urlprefix\endcsname\relax\def\urlprefix{URL }\fi
\providecommand{\bibinfo}[2]{#2}
\providecommand{\eprint}[2][]{\url{#2}}

\bibitem[{\citenamefont{Emmerich et~al.}(2012)\citenamefont{Emmerich,
  L{\"o}wen, Wittkowski, Gruhn, T{\'o}th, Tegze, and
  Gr{\'a}n{\'a}sy}}]{Emmerich2012}
\bibinfo{author}{\bibfnamefont{H.}~\bibnamefont{Emmerich}},
  \bibinfo{author}{\bibfnamefont{H.}~\bibnamefont{L{\"o}wen}},
  \bibinfo{author}{\bibfnamefont{R.}~\bibnamefont{Wittkowski}},
  \bibinfo{author}{\bibfnamefont{T.}~\bibnamefont{Gruhn}},
  \bibinfo{author}{\bibfnamefont{G.~I.} \bibnamefont{T{\'o}th}},
  \bibinfo{author}{\bibfnamefont{G.}~\bibnamefont{Tegze}}, \bibnamefont{and}
  \bibinfo{author}{\bibfnamefont{L.}~\bibnamefont{Gr{\'a}n{\'a}sy}},
  \bibinfo{journal}{Adv. Phys.} \textbf{\bibinfo{volume}{61}},
  \bibinfo{pages}{665} (\bibinfo{year}{2012}),
  \urlprefix\url{http://dx.doi.org/10.1080/00018732.2012.737555}.

\bibitem[{\citenamefont{van Teeffelen et~al.}(2009)\citenamefont{van Teeffelen,
  Backofen, Voigt, and L{\"o}wen}}]{van2009derivation}
\bibinfo{author}{\bibfnamefont{S.}~\bibnamefont{van Teeffelen}},
  \bibinfo{author}{\bibfnamefont{R.}~\bibnamefont{Backofen}},
  \bibinfo{author}{\bibfnamefont{A.}~\bibnamefont{Voigt}}, \bibnamefont{and}
  \bibinfo{author}{\bibfnamefont{H.}~\bibnamefont{L{\"o}wen}},
  \bibinfo{journal}{Phys. Rev. E} \textbf{\bibinfo{volume}{79}},
  \bibinfo{pages}{051404} (\bibinfo{year}{2009}),
  \urlprefix\url{https://journals.aps.org/pre/abstract/10.1103/PhysRevE.79.051%
404}.

\bibitem[{\citenamefont{Archer et~al.}(2012)\citenamefont{Archer, Robbins,
  Thiele, and Knobloch}}]{archer2012solidification}
\bibinfo{author}{\bibfnamefont{A.~J.} \bibnamefont{Archer}},
  \bibinfo{author}{\bibfnamefont{M.~J.} \bibnamefont{Robbins}},
  \bibinfo{author}{\bibfnamefont{U.}~\bibnamefont{Thiele}}, \bibnamefont{and}
  \bibinfo{author}{\bibfnamefont{E.}~\bibnamefont{Knobloch}},
  \bibinfo{journal}{Phys. Rev. E} \textbf{\bibinfo{volume}{86}},
  \bibinfo{pages}{031603} (\bibinfo{year}{2012}),
  \urlprefix\url{https://journals.aps.org/pre/abstract/10.1103/PhysRevE.86.031%
603}.

\bibitem[{\citenamefont{Evans}(1992)}]{Evans1992}
\bibinfo{author}{\bibfnamefont{R.}~\bibnamefont{Evans}}, in
  \emph{\bibinfo{booktitle}{Fundamentals of Inhomogenous Fluids}}, edited by
  \bibinfo{editor}{\bibfnamefont{D.}~\bibnamefont{Henderson}}
  (\bibinfo{publisher}{Marcel Decker}, \bibinfo{address}{New York},
  \bibinfo{year}{1992}), chap.~\bibinfo{chapter}{3}, pp.
  \bibinfo{pages}{85--176},
  \bibinfo{note}{\url{https://www.crcpress.com/Fundamentals-of-Inhomogeneous-F%
luids/Henderson/p/book/9780824787110}}.

\bibitem[{\citenamefont{Hansen and McDonald}(2013)}]{Hansen2013}
\bibinfo{author}{\bibfnamefont{J.-P.} \bibnamefont{Hansen}} \bibnamefont{and}
  \bibinfo{author}{\bibfnamefont{I.~R.} \bibnamefont{McDonald}},
  \emph{\bibinfo{title}{Theory of Simple Liquids with Applications to Soft
  Matter: Fourth Edition}} (\bibinfo{publisher}{Elsevier},
  \bibinfo{address}{Oxford}, \bibinfo{year}{2013}),
  \urlprefix\url{https://www.elsevier.com/books/theory-of-simple-liquids/hanse%
n/978-0-12-387032-2}.

\bibitem[{\citenamefont{Thiele et~al.}(2013)\citenamefont{Thiele, Archer,
  Robbins, Gomez, and Knobloch}}]{Thiele2013}
\bibinfo{author}{\bibfnamefont{U.}~\bibnamefont{Thiele}},
  \bibinfo{author}{\bibfnamefont{A.~J.} \bibnamefont{Archer}},
  \bibinfo{author}{\bibfnamefont{M.~J.} \bibnamefont{Robbins}},
  \bibinfo{author}{\bibfnamefont{H.}~\bibnamefont{Gomez}}, \bibnamefont{and}
  \bibinfo{author}{\bibfnamefont{E.}~\bibnamefont{Knobloch}},
  \bibinfo{journal}{Phys. Rev. E} \textbf{\bibinfo{volume}{87}},
  \bibinfo{pages}{042915} (\bibinfo{year}{2013}),
  \urlprefix\url{https://journals.aps.org/pre/abstract/10.1103/PhysRevE.87.042%
915}.

\bibitem[{\citenamefont{Elder and Grant}(2004)}]{elder2004modeling}
\bibinfo{author}{\bibfnamefont{K.~R.} \bibnamefont{Elder}} \bibnamefont{and}
  \bibinfo{author}{\bibfnamefont{M.}~\bibnamefont{Grant}},
  \bibinfo{journal}{Phys. Rev. E} \textbf{\bibinfo{volume}{70}},
  \bibinfo{pages}{051605} (\bibinfo{year}{2004}),
  \urlprefix\url{https://journals.aps.org/pre/abstract/10.1103/PhysRevE.70.051%
605}.

\bibitem[{\citenamefont{Knobloch}(2016)}]{Knobloch2016}
\bibinfo{author}{\bibfnamefont{E.}~\bibnamefont{Knobloch}},
  \bibinfo{journal}{IMA Journal of Applied Mathematics}
  \textbf{\bibinfo{volume}{81}}, \bibinfo{pages}{457} (\bibinfo{year}{2016}),
  \urlprefix\url{https://academic.oup.com/imamat/article/81/3/457/2871034/Loca%
lized-structures-and-front-propagation-in}.

\bibitem[{\citenamefont{Engel and Trebin}(2007)}]{Engel2007}
\bibinfo{author}{\bibfnamefont{M.}~\bibnamefont{Engel}} \bibnamefont{and}
  \bibinfo{author}{\bibfnamefont{H.-R.} \bibnamefont{Trebin}},
  \bibinfo{journal}{Phys. Rev. Lett.} \textbf{\bibinfo{volume}{98}},
  \bibinfo{pages}{225505} (\bibinfo{year}{2007}),
  \urlprefix\url{https://journals.aps.org/prl/abstract/10.1103/PhysRevLett.98.%
225505}.

\bibitem[{\citenamefont{Subramanian et~al.}(2016)\citenamefont{Subramanian,
  Archer, Knobloch, and Rucklidge}}]{Subramanian2016}
\bibinfo{author}{\bibfnamefont{P.}~\bibnamefont{Subramanian}},
  \bibinfo{author}{\bibfnamefont{A.~J.} \bibnamefont{Archer}},
  \bibinfo{author}{\bibfnamefont{E.}~\bibnamefont{Knobloch}}, \bibnamefont{and}
  \bibinfo{author}{\bibfnamefont{A.~M.} \bibnamefont{Rucklidge}},
  \bibinfo{journal}{Phys. Rev. Lett.} \textbf{\bibinfo{volume}{117}},
  \bibinfo{pages}{075501} (\bibinfo{year}{2016}),
  \urlprefix\url{https://journals.aps.org/prl/abstract/10.1103/PhysRevLett.117%
.075501}.

\bibitem[{\citenamefont{Achim et~al.}(2014)\citenamefont{Achim, Schmiedeberg,
  and L\"owen}}]{Achim2014}
\bibinfo{author}{\bibfnamefont{C.~V.} \bibnamefont{Achim}},
  \bibinfo{author}{\bibfnamefont{M.}~\bibnamefont{Schmiedeberg}},
  \bibnamefont{and} \bibinfo{author}{\bibfnamefont{H.}~\bibnamefont{L\"owen}},
  \bibinfo{journal}{Phys. Rev. Lett.} \textbf{\bibinfo{volume}{112}},
  \bibinfo{pages}{255501} (\bibinfo{year}{2014}),
  \urlprefix\url{https://link.aps.org/doi/10.1103/PhysRevLett.112.255501}.

\bibitem[{\citenamefont{Rucklidge et~al.}(2012)\citenamefont{Rucklidge, Silber,
  and Skeldon}}]{RucklidgePRL2012}
\bibinfo{author}{\bibfnamefont{A.~M.} \bibnamefont{Rucklidge}},
  \bibinfo{author}{\bibfnamefont{M.}~\bibnamefont{Silber}}, \bibnamefont{and}
  \bibinfo{author}{\bibfnamefont{A.~C.} \bibnamefont{Skeldon}},
  \bibinfo{journal}{Phys. Rev. Lett.} \textbf{\bibinfo{volume}{108}},
  \bibinfo{pages}{074504} (\bibinfo{year}{2012}),
  \urlprefix\url{http://link.aps.org/doi/10.1103/PhysRevLett.108.074504}.

\bibitem[{\citenamefont{Lifshitz and Petrich}(1997)}]{Lifshitz1997}
\bibinfo{author}{\bibfnamefont{R.}~\bibnamefont{Lifshitz}} \bibnamefont{and}
  \bibinfo{author}{\bibfnamefont{D.~M.} \bibnamefont{Petrich}},
  \bibinfo{journal}{Phys. Rev. Lett.} \textbf{\bibinfo{volume}{79}},
  \bibinfo{pages}{1261} (\bibinfo{year}{1997}),
  \urlprefix\url{http://link.aps.org/doi/10.1103/PhysRevLett.79.1261}.

\bibitem[{\citenamefont{Archer et~al.}(2015)\citenamefont{Archer, Rucklidge,
  and Knobloch}}]{Archer2015}
\bibinfo{author}{\bibfnamefont{A.~J.} \bibnamefont{Archer}},
  \bibinfo{author}{\bibfnamefont{A.~M.} \bibnamefont{Rucklidge}},
  \bibnamefont{and} \bibinfo{author}{\bibfnamefont{E.}~\bibnamefont{Knobloch}},
  \bibinfo{journal}{Phys. Rev. E} \textbf{\bibinfo{volume}{92}},
  \bibinfo{pages}{012324} (\bibinfo{year}{2015}),
  \urlprefix\url{http://link.aps.org/doi/10.1103/PhysRevE.92.012324}.

\bibitem[{\citenamefont{Subramanian et~al.}(2017)\citenamefont{Subramanian,
  Archer, Knobloch, and Rucklidge}}]{Supplement_SLQ}
\bibinfo{author}{\bibfnamefont{P.}~\bibnamefont{Subramanian}},
  \bibinfo{author}{\bibfnamefont{A.~J.} \bibnamefont{Archer}},
  \bibinfo{author}{\bibfnamefont{E.}~\bibnamefont{Knobloch}}, \bibnamefont{and}
  \bibinfo{author}{\bibfnamefont{A.~M.} \bibnamefont{Rucklidge}},
  \emph{\bibinfo{title}{Supplementary video detailing structure of an extended
  and spatially localized $3d$~phase field quasicrystals}}
  (\bibinfo{year}{2017}).

\bibitem[{\citenamefont{Rucklidge and Silber}(2009)}]{Rucklidge2009}
\bibinfo{author}{\bibfnamefont{A.~M.} \bibnamefont{Rucklidge}}
  \bibnamefont{and} \bibinfo{author}{\bibfnamefont{M.}~\bibnamefont{Silber}},
  \bibinfo{journal}{SIAM J. Appl. Dyn. Sys.} \textbf{\bibinfo{volume}{8}},
  \bibinfo{pages}{298} (\bibinfo{year}{2009}),
  \urlprefix\url{http://epubs.siam.org/doi/10.1137/080719066}.

\bibitem[{\citenamefont{Frigo and Johnson}(2005)}]{FFTW05}
\bibinfo{author}{\bibfnamefont{M.}~\bibnamefont{Frigo}} \bibnamefont{and}
  \bibinfo{author}{\bibfnamefont{S.~G.} \bibnamefont{Johnson}},
  \bibinfo{journal}{Proceedings of the IEEE} \textbf{\bibinfo{volume}{93}},
  \bibinfo{pages}{216} (\bibinfo{year}{2005}), \bibinfo{note}{special issue on
  ``Program Generation, Optimization, and Platform Adaptation''}.

\bibitem[{\citenamefont{Cox and Matthews}(2002)}]{CoxJCP2002}
\bibinfo{author}{\bibfnamefont{S.~M.} \bibnamefont{Cox}} \bibnamefont{and}
  \bibinfo{author}{\bibfnamefont{P.~C.} \bibnamefont{Matthews}},
  \bibinfo{journal}{Journal of Computational Physics}
  \textbf{\bibinfo{volume}{176}}, \bibinfo{pages}{430 } (\bibinfo{year}{2002}),
  ISSN \bibinfo{issn}{0021-9991},
  \urlprefix\url{http://www.sciencedirect.com/science/article/pii/S00219991029%
69950}.

\bibitem[{\citenamefont{Doedel et~al.}(1991)\citenamefont{Doedel, Keller, and
  Kernevez}}]{Doedel1991}
\bibinfo{author}{\bibfnamefont{E.}~\bibnamefont{Doedel}},
  \bibinfo{author}{\bibfnamefont{H.~B.} \bibnamefont{Keller}},
  \bibnamefont{and} \bibinfo{author}{\bibfnamefont{J.~P.}
  \bibnamefont{Kernevez}}, \bibinfo{journal}{Int. J. Bifurcation and Chaos}
  \textbf{\bibinfo{volume}{1}}, \bibinfo{pages}{493} (\bibinfo{year}{1991}),
  \urlprefix\url{http://www.worldscientific.com/doi/abs/10.1142/S0218127491000%
397}.

\bibitem[{\citenamefont{van~der Vorst}(1992)}]{vanderVorst1992}
\bibinfo{author}{\bibfnamefont{H.~A.} \bibnamefont{van~der Vorst}},
  \bibinfo{journal}{SIAM Journal on Scientific and Statistical Computing}
  \textbf{\bibinfo{volume}{13}}, \bibinfo{pages}{631} (\bibinfo{year}{1992}),
  \urlprefix\url{http://epubs.siam.org/doi/abs/10.1137/0913035}.

\bibitem[{\citenamefont{Sonneveld and van Gijzen}(2008)}]{Sonneveld2008}
\bibinfo{author}{\bibfnamefont{P.}~\bibnamefont{Sonneveld}} \bibnamefont{and}
  \bibinfo{author}{\bibfnamefont{M.~B.} \bibnamefont{van Gijzen}},
  \bibinfo{journal}{SIAM Journal on Scientific Computing}
  \textbf{\bibinfo{volume}{31}}, \bibinfo{pages}{1035} (\bibinfo{year}{2008}),
  \urlprefix\url{http://epubs.siam.org/doi/abs/10.1137/070685804}.

\bibitem[{\citenamefont{Bergeon et~al.}(2008)\citenamefont{Bergeon, Burke,
  Knobloch, and Mercader}}]{Bergeon2008}
\bibinfo{author}{\bibfnamefont{A.}~\bibnamefont{Bergeon}},
  \bibinfo{author}{\bibfnamefont{J.}~\bibnamefont{Burke}},
  \bibinfo{author}{\bibfnamefont{E.}~\bibnamefont{Knobloch}}, \bibnamefont{and}
  \bibinfo{author}{\bibfnamefont{I.}~\bibnamefont{Mercader}},
  \bibinfo{journal}{Phys. Rev. E} \textbf{\bibinfo{volume}{78}},
  \bibinfo{pages}{046201} (\bibinfo{year}{2008}),
  \urlprefix\url{https://link.aps.org/doi/10.1103/PhysRevE.78.046201}.

\bibitem[{\citenamefont{Dawes}(2008)}]{Dawes2008}
\bibinfo{author}{\bibfnamefont{J.~H.~P.} \bibnamefont{Dawes}},
  \bibinfo{journal}{SIAM J. Appl. Dyn. Sys.} \textbf{\bibinfo{volume}{7}},
  \bibinfo{pages}{186} (\bibinfo{year}{2008}),
  \eprint{https://doi.org/10.1137/06067794X},
  \urlprefix\url{https://doi.org/10.1137/06067794X}.

\bibitem[{\citenamefont{Lloyd et~al.}(2008)\citenamefont{Lloyd, Sandstede,
  Avitabile, and Champneys}}]{Lloyd2008}
\bibinfo{author}{\bibfnamefont{D.~J.~B.} \bibnamefont{Lloyd}},
  \bibinfo{author}{\bibfnamefont{B.}~\bibnamefont{Sandstede}},
  \bibinfo{author}{\bibfnamefont{D.}~\bibnamefont{Avitabile}},
  \bibnamefont{and} \bibinfo{author}{\bibfnamefont{A.~R.}
  \bibnamefont{Champneys}}, \bibinfo{journal}{SIAM J. Appl. Dyn. Sys.}
  \textbf{\bibinfo{volume}{7}}, \bibinfo{pages}{1049} (\bibinfo{year}{2008}),
  \eprint{https://doi.org/10.1137/070707622},
  \urlprefix\url{https://doi.org/10.1137/070707622}.

\bibitem[{\citenamefont{Barkan et~al.}(2014)\citenamefont{Barkan, Engel, and
  Lifshitz}}]{Barkan2014}
\bibinfo{author}{\bibfnamefont{K.}~\bibnamefont{Barkan}},
  \bibinfo{author}{\bibfnamefont{M.}~\bibnamefont{Engel}}, \bibnamefont{and}
  \bibinfo{author}{\bibfnamefont{R.}~\bibnamefont{Lifshitz}},
  \bibinfo{journal}{Phys. Rev. Lett.} \textbf{\bibinfo{volume}{113}},
  \bibinfo{pages}{098304} (\bibinfo{year}{2014}),
  \urlprefix\url{https://link.aps.org/doi/10.1103/PhysRevLett.113.098304}.

\bibitem[{\citenamefont{Tanaka}(1999)}]{tanaka1999two}
\bibinfo{author}{\bibfnamefont{H.}~\bibnamefont{Tanaka}}, \bibinfo{journal}{J.
  Chem. Phys.} \textbf{\bibinfo{volume}{111}}, \bibinfo{pages}{3163}
  (\bibinfo{year}{1999}),
  \urlprefix\url{http://aip.scitation.org/doi/abs/10.1063/1.479596}.

\bibitem[{\citenamefont{Mossa and Tarjus}(2003)}]{mossa2003locally}
\bibinfo{author}{\bibfnamefont{S.}~\bibnamefont{Mossa}} \bibnamefont{and}
  \bibinfo{author}{\bibfnamefont{G.}~\bibnamefont{Tarjus}},
  \bibinfo{journal}{J. Chem. Phys.} \textbf{\bibinfo{volume}{119}},
  \bibinfo{pages}{8069} (\bibinfo{year}{2003}),
  \urlprefix\url{http://aip.scitation.org/doi/abs/10.1063/1.1604380}.

\bibitem[{\citenamefont{Royall et~al.}(2008)\citenamefont{Royall, Williams,
  Ohtsuka, and Tanaka}}]{Royall2008}
\bibinfo{author}{\bibfnamefont{C.~P.} \bibnamefont{Royall}},
  \bibinfo{author}{\bibfnamefont{S.~R.} \bibnamefont{Williams}},
  \bibinfo{author}{\bibfnamefont{T.}~\bibnamefont{Ohtsuka}}, \bibnamefont{and}
  \bibinfo{author}{\bibfnamefont{H.}~\bibnamefont{Tanaka}},
  \bibinfo{journal}{Nature Materials} \textbf{\bibinfo{volume}{7}},
  \bibinfo{pages}{556} (\bibinfo{year}{2008}),
  \urlprefix\url{http://www.nature.com/nmat/journal/v7/n7/full/nmat2219.html?f%
oxtrotcallback=true}.

\end{thebibliography}

\end{document}